\documentclass[12pt,titlepage]{article}
\usepackage{amssymb}

\input{tcilatex}

\begin{document}

\title{Trapping neutral fermions \\
with kink-like potentials}
\date{}
\author{Antonio S. de Castro$^{a,b}$ and Marcelo Hott$^{b,}$\thanks{%
E-mail address: hott@feg.unesp.br (M.Hott)} \\
\\
$^{a}$Universidade de Coimbra\\
Centro de F\'{\i}sica Computacional\\
P-3004-516 Coimbra - Portugal\\
\smallskip \\
$^{b}$UNESP - Campus de Guaratinguet\'{a}\\
Departamento de F\'{\i}sica e Qu\'{\i}mica\\
12516-410 Guaratinguet\'{a} SP - Brasil\\
\\
}
\maketitle

\begin{abstract}
The intrinsically relativistic problem of neutral fermions subject to
kink--like potentials ($\sim \mathrm{tanh}\,\gamma x$) is investigated and
the exact bound-state solutions are found. Apart from the lonely hump
solutions for $E=\pm mc^{2}$, the problem is mapped into the exactly
solvable Surm-Liouville problem with a modified P\"{o}schl-Teller potential.
An apparent paradox concerning the uncertainty principle is solved by
resorting to the concepts of effective mass and effective Compton wavelength.
\end{abstract}

\section{Introduction}

The four-dimensional Dirac equation with an anomalous magnetic-like (tensor)
coupling describes the interaction of neutral fermions with electric fields
and can be reduced to the two-dimensional Dirac equation with a pseudoscalar
coupling when the fermion is limited to move in just one direction.
Therefore, the investigation of the simpler Dirac equation in a 1+1
dimension with a pseudoscalar potential might be relevant to a better
understanding of the problem of neutral fermions subject to electric fields
in the more realistic 3+1 world.

The states of fermions in one-plus-one dimensions bound by a pseudoscalar
double-step potential \cite{asc2} and their scattering by a pseudoscalar
step potential \cite{asc3} have already been analyzed and some quite
interesting results have been found. Indeed, the two-dimensional version of
the anomalous magnetic-like interaction linear in the radial coordinate,
christened by Moshinsky and Szczepaniak \cite{ms} as Dirac oscillator and
extensively studied before \cite{ito}-\cite{hug}, has also received
attention. Nogami and Toyama \cite{nt}, Toyama et al. \cite{tplus} and
Toyama and Nogami \cite{tn} studied the behaviour of wave packets under the
influence of that parity-conserving potential whereas Szmytkowski and
Gruchowski \cite{sg} proved the completeness of the eigenfunctions. More
recently Pacheco et al. \cite{pa} studied a few thermodynamic properties of
the 1+1 dimensional Dirac oscillator, and a generalization of the Dirac
oscillator for a negative coupling constant was presented in Ref. \cite{asc}%
. The two-dimensional generalized Dirac oscillator plus an inversely linear
potential has also been addressed in Ref. \cite{asc4}.

In recent papers, Villalba \cite{vil} and McKeon and Van Leeuwen \cite{mck}
considered a pseu\-do\-sca\-lar Coulomb potential ($V=\lambda /r$) in 3+1
dimensions and concluded that there are no bound states. The reason
attributed in Ref. \cite{mck} for the absence of bound-state solutions is
that the different parity eigenstates mix. Furthermore, the authors of Ref. 
\cite{mck} assert that \textit{the absence of bound states in this system
confuses the role of the }$\pi $\textit{-meson in the binding of nucleons}.
Such an intriguing conclusion sets the stage for the analysis by other sorts
of pseudoscalar potentials. A natural question to ask is whether the absence
of bound-state solutions by a pseudoscalar Coulomb potential is a
characteristic feature of the four-dimensional world. In Ref. \cite{asc} the
Dirac equation in one-plus-one dimensions with the pseudoscalar power-law
potential $V=\mu |x|^{\delta }$ was approached and there it was concluded
that $V$ is a binding potential only for $\delta >0$. That conclusion
sharply contrasts with the result found in \cite{mck}. Ref. \cite{asc} shows
that it is possible to find bound states for fermions interacting by a
pseudoscalar potential in 1+1 dimensions despite the fact that the spinor is
not an eigenfunction of the parity operator.

The parity-conserving pseudoscalar potential $\sim \mathrm{tanh}\,\gamma x$
is of interest in quantum field theory where topological classical
backgrounds are responsible for inducing a fractional fermion number on the
vacuum. Models of this kind, known as kink models, are obtained in quantum
field theory as the continuum limit of linear polymer models \cite{gol}-\cite%
{sem}. To the best of our knowledge, no one has computed the complete set of
bound states in the presence of this sort of potential. The present work
investigates the bound-state solutions of fermions immersed in the
background of the pseudoscalar potential $V=\hbar c\gamma g\,\mathrm{tanh}%
\,\gamma x$, termed kink-like potential$.$ A peculiar feature of this
potential is the absence of bound states in a nonrelativistic theory because
it gives rise to an ubiquitous repulsive potential. The whole spectrum of
this intrinsically relativistic problem is found analytically, for both
massive fermions and massless fermions. Fortunately, apart from solutions
corresponding to $|E|=mc^{2}$, the problem is reducible to the finite set of
solutions of the nonrelativistic exactly solvable symmetric modified P\"{o}%
schl-Teller potential for both components of the Dirac spinor subject to a
constraint on their nodal structure. Finally, we observe a remarkable
feature of this problem: the possibility of trapping a fermion with an
uncertainty in the position that can shrink without limit as $|\gamma |$ and 
$|g|$ increase without violating the Heisenberg uncertainty principle.

\section{The Dirac equation with a pseudoscalar potential in a 1+1 dimension}

The 1+1 dimensional time-independent Dirac equation for a fermion of rest
mass $m$ coupled to a pseudoscalar potential reads

\begin{equation}
H\psi =E\psi ,\quad H=c\alpha p+\beta mc^{2}+\beta \gamma ^{5}V  \label{1}
\end{equation}

\noindent where $E$ is the energy of the fermion, $c$ is the velocity of
light and $p$ is the momentum operator. The positive definite function $%
|\psi |^{2}=\psi ^{\dagger }\psi $, satisfying a continuity equation, is
interpreted as a position probability density and its norm is a constant of
motion. This interpretation is completely satisfactory for single-particle
states \cite{tha}. We use $\alpha =\sigma _{1}$ and $\beta =\sigma _{3}$,
where $\sigma _{1}$ and $\sigma _{3}$ are Pauli matrices, and $\beta \gamma
^{5}=\sigma _{2}$. Provided that the spinor is written in terms of the upper
and the lower components, $\psi _{+}$ and $\psi _{-}$ respectively,
\noindent the Dirac equation decomposes into:

\begin{equation}
\left( -E\pm mc^{2}\right) \psi _{\pm }=i\hbar c\psi _{\mp }^{\prime }\pm
iV\psi _{\mp }  \label{2}
\end{equation}

\noindent where the prime denotes differentiation with respect to $x$. In
terms of $\psi _{+}$ and $\psi _{-}$ the spinor is normalized as $%
\int_{-\infty }^{+\infty }dx\left( |\psi _{+}|^{2}+|\psi _{-}|^{2}\right) =1$
so that $\psi _{+}$ and $\psi _{-}$ are square integrable functions. It is
clear from the pair of coupled first-order differential equations given by (%
\ref{2}) that $\psi _{+}$ and $\psi _{-}$ have definite and opposite
parities if the Dirac equation is covariant under $x\rightarrow -x$, i.e. if
the pseudoscalar potential function is odd. The charge conjugation operation
requires that if $\psi $ is a solution with eigenenergy $E$ for the
potential $V$ then $\sigma _{1}\psi ^{\ast }$ is a solution with eigenenergy 
$-E$ for the potential $-V$. It is interesting to note that the operation of
just interchanging the upper and lower components of the Dirac spinor
induced by $i\gamma ^{5}\psi $ preserves the eigenenergies for a massless
fermion when $V\rightarrow -V$. One can also see that the operator $\mathcal{%
O}=i\left[ H,\sigma _{3}\right] /2$ anticommutes with $H$ so that it maps
positive- into negative-energy solutions, and vice versa. Although this last
operator does not preserve the norm for scattering states, it can be used to
obtain the normalized states corresponding to eigenenergies $-E$ from the
knowledge of the normalized states with eigenenergies $E.$

In the nonrelativistic approximation (potential energies small compared to $%
mc^{2}$ and $E\approx mc^{2}$) Eq. (\ref{2}) becomes

\begin{equation}
\psi _{-}=\left( \frac{p}{2mc}\,+i\,\frac{V}{2mc^{2}}\right) \psi _{+}
\label{3}
\end{equation}

\begin{equation}
\left( -\frac{\hbar ^{2}}{2m}\frac{d^{2}}{dx^{2}}+\frac{V^{2}}{2mc^{2}}+%
\frac{\hbar V^{\prime }}{2mc}\right) \psi _{+}=\left( E-mc^{2}\right) \psi
_{+}  \label{4}
\end{equation}

\noindent Eq. (\ref{3}) shows that $\psi _{-}$ is of order $v/c<<1$ relative
to $\psi _{+}$ and Eq. (\ref{4}) shows that $\psi _{+}$ obeys the Schr\"{o}%
dinger equation. Note that the pseudoscalar coupling has the effect that the
Schr\"{o}dinger equation has an effective potential in the nonrelativistic
limit, and not the original potential itself. Indeed, this is the same side
effect which in a 3+1 dimensional space-time makes the tensor linear
potential to manifest itself as a harmonic oscillator plus a strong
spin-orbit coupling in the nonrelativistic limit \cite{ms}. The form in
which the original potential appears in the effective potential, the $V^{2}$
term, allows us to infer that even a potential unbounded from below could be
a confining potential. This phenomenon is inconceivable if one starts with
the original potential in the nonrelativistic equation.

It should be noted that $V\rightarrow V+$const in the Dirac equation and in
its nonrelativistic limit does not yield $E\rightarrow E+$ const. Therefore,
the potential and the energy themselves \ and not just the potential and
energy differences have physical significance. It has already been verified
that a constant added to the screened Coulomb potential \cite{asc10} or to
the inversely linear potential \cite{asc100} is undoubtedly physically
relevant. As a matter of fact, it plays a crucial role in ensuring the
existence of bound states.

For $E\neq \pm mc^{2}$, the coupling between the upper and the lower
components of the Dirac spinor can be formally eliminated when Eqs. (\ref{2}%
) are written as second-order differential equations:

\begin{equation}
-\frac{\hbar ^{2}}{2}\,\psi _{\pm }^{\prime \prime }+\left( \frac{V^{2}}{%
2c^{2}}\pm \frac{\hbar }{2c}V^{\prime }\right) \,\psi _{\pm }=\frac{%
E^{2}-m^{2}c^{4}}{2c^{2}}\,\psi _{\pm }  \label{5}
\end{equation}

\noindent This last result shows that the solution for this class of problem
consists in searching for bound-state solutions for two Schr\"{o}dinger
equations. It should not be forgotten, though, that the equations for $\psi
_{+}$ or $\psi _{-}$ are not indeed independent because $E$ appears in both
equations. Therefore, one has to search for bound-state solutions for both
signs in (\ref{5}) with a common eigenvalue. At this stage one can realize
that \noindent \noindent the Dirac energy levels are symmetrical about $E=0$%
. This means that the potential couples to the positive-energy component of
the spinor in the same way it couples to the negative-energy component. In
other words, this sort of potential couples to the mass of the fermion
instead of its charge, so that there is no atmosphere for the spontaneous
production of particle-antiparticle pairs. No matter what the intensity and
sign of the potential is, the positive- and the negative-energy solutions
never meet each other. Thus there is no room for transitions from positive-
to negative-energy solutions. This all means that Klein\'{}s paradox never
comes into the scenario.

The solutions for $E=\pm mc^{2}$, excluded from the Sturm-Liouville problem,
can be obtained directly from the Dirac equation (\ref{2}). One can observe
that such isolated solutions, for $E=+mc^{2}$, are

\begin{eqnarray}
\psi _{-} &=&N_{-}\,\exp \left[ -v(x)\right]  \nonumber \\
&&  \label{6} \\
\psi _{+}^{\prime }-v^{\prime }\psi _{+} &=&+i\,\frac{2mc}{\hbar }%
N_{-}\,\exp \left[ -v(x)\right]  \nonumber
\end{eqnarray}

\noindent and, for $E=-mc^{2}$,

\begin{eqnarray}
\psi _{+} &=&N_{+}\,\exp \left[ +v(x)\right]  \nonumber \\
&&  \label{6a} \\
\psi _{-}^{\prime }+v^{\prime }\psi _{-} &=&-i\,\frac{2mc}{\hbar }%
N_{+}\,\exp \left[ +v(x)\right]  \nonumber
\end{eqnarray}

\noindent where $N_{+}$ and $N_{-}$ are normalization constants and $%
v(x)=\int^{x}dy\,V(y)\,/(\hbar c)$. \noindent Of course well-behaved
eigenstates are possible only if $v(x)$ has an appropriate leading
asymptotic behaviour.

\section{The kink-like potential}

Now let us concentrate our attention on the potential 
\begin{equation}
V=\hbar c\gamma g\,\mathrm{tanh}\,\gamma x  \label{7}
\end{equation}
\noindent where $\gamma $ and the dimensionless coupling constant, $g$, are
real numbers. The potential is invariant under the change $\gamma
\rightarrow -\gamma $ so that the results can depend only on $|\gamma |$
whereas the sign of $V$ depends on the sign of $g$. Since the solutions for
different signs of $g$ can be connected by the charge conjugation
transformation, and by the chiral transformation in the event of massless
fermions, we restrict ourselves to the case $g>0$.

The Sturm-Liouville problem corresponding to Eq. (\ref{5}) becomes

\begin{equation}
-\frac{\hbar ^{2}}{2m_{eff}}\,\psi _{\pm }^{\prime \prime }+V_{eff}^{[\pm
]}\,\psi _{\pm }=E_{eff}\,\psi _{\pm }  \label{8}
\end{equation}

\noindent where we recognize the effective potential as the exactly solvable
symmetric modified P\"{o}schl-Teller potential \cite{rm}-\cite{flu} (in the
notation of Refs. \cite{lan} and \cite{nie})

\begin{equation}
V_{eff}^{[\pm ]}(x)=-U_{0}^{[\pm ]}\,\mathrm{sech}^{2}\gamma x,\quad
U_{0}^{[\pm ]}=\frac{\hbar ^{2}\gamma ^{2}}{2m_{eff}}\,g\left( g\mp 1\right)
>0\Rightarrow g>1  \label{9}
\end{equation}%
whose normalizable eigenfunctions corresponding to bound-state solutions,
subject to the boundary conditions $\psi _{\pm }=0$ as $|x|\rightarrow
\infty $, are possible only if the effective potentials for both $\psi _{+}$
and $\psi _{-}$ present potential-well structures. According to (\ref{9}),
this demands that $g>1$. The corresponding effective eigenenergy is given by

\noindent

\begin{equation}
E_{eff}=\frac{E^{2}-m_{eff}^{2}c^{4}}{2m_{eff}c^{2}}=-\,\frac{\hbar
^{2}\gamma ^{2}}{2m_{eff}}\left( s_{\pm }-n_{\pm }\right) ^{2}  \label{10}
\end{equation}

\noindent where

\begin{equation}
s_{\pm }=\frac{1}{2}\left( -1+\sqrt{1+\frac{8m_{eff}U_{0}^{[\pm ]}}{\hbar
^{2}\gamma ^{2}}}\,\right) \Rightarrow \left\{ 
\begin{array}{c}
s_{+} \\ 
s_{-}%
\end{array}
\begin{array}{l}
=g-1 \\ 
=g%
\end{array}
\right.  \label{10a}
\end{equation}
\begin{equation}
n_{\pm }=0,1,2,\ldots <s_{\pm }  \label{10a1}
\end{equation}

\begin{equation}
m_{eff}=\sqrt{m^{2}+\left( \frac{\hbar \gamma g}{c}\right) ^{2}}  \label{10c}
\end{equation}

\noindent Notice that $V_{eff}^{[\pm ]}$ is an even function under $%
x\rightarrow -x$. Furthermore, Eqs. (\ref{10a}) and (\ref{10a1}) show that
the capacity of the potential to hold bound-state solutions is independent
of $\gamma $. As for $g$, it can be seen that the number of allowed bound
states depends linearly on $g$ and there is always at least one bound-state
solution for any $g>1$. From (\ref{9}) and (\ref{10}) one can note that the
Dirac eigenenergies related to the bound-state solutions are restricted to
the range 
\begin{equation}
\sqrt{m^{2}c^{4}+\left( \hbar c\gamma \right) ^{2}g}<|E|<\sqrt{%
m^{2}c^{4}+\left( \hbar c\gamma \right) ^{2}g^{2}}  \label{100}
\end{equation}%
and that the eigenenergies in the range $|E|>\sqrt{m^{2}c^{4}+\left( \hbar
c\gamma \right) ^{2}g^{2}}$ correspond to the continuum. Since the positive-
and negative-eigenenergies never intercept each other, one can see once
again that Klein\'{}s paradox is absent from this picture. In order to match
the common effective eigenvalue for the effective potentials $V_{eff}^{[+]}$
and $V_{eff}^{[-]}$ one can see from (\ref{10a}) and (\ref{10a1}) that the
following constraint 
\begin{equation}
a_{n}=s_{+}-n_{+}=s_{-}-n_{-}=g-1-n_{+}  \label{10d}
\end{equation}%
\noindent must be satisfied. Eq. (\ref{10d}) implies that the quantum
numbers $n_{+}$ and $n_{-}$ satisfy the relation 
\begin{equation}
n_{-}=n_{+}+1  \label{10f}
\end{equation}%
This last fact can be better understood by observing that $V_{eff}^{[+]}$ is
deeper than $V_{eff}^{[-]}$. Now, (\ref{10})-(\ref{10c}) tell us that

\begin{equation}
E=\pm \,\sqrt{m^{2}c^{4}+\left( \hbar c\gamma \right) ^{2}\left(
g^{2}-a_{n}^{2}\right) }  \label{10e}
\end{equation}
where 
\[
n_{+}=0,1,2,\ldots <g-1 
\]
The upper and lower components of the Dirac spinor can be written as (see
Ref. \cite{nie})

\begin{equation}
\psi _{\pm }=N_{\pm }\,2^{a_{n}}\Gamma \left( a_{n}+\frac{1}{2}\right) \sqrt{%
\frac{|\gamma |a_{n}}{\pi }\frac{\Gamma \left( n_{_{\pm }}+1\right) }{\Gamma
\left( n_{_{\pm }}+1+2a_{n}\right) }}\,\left( 1-z^{2}\right)
^{a_{n}/2}C_{n_{_{\pm }}}^{\left( a_{n}+1/2\right) }\left( z\right)
\label{21}
\end{equation}

\noindent where $z=\mathrm{tanh}\,\gamma x$ and $C_{n}^{\left( a\right)
}\left( z\right) $ is the Gegenbauer (ultraspherical) polynomial of degree $%
n $. Since $C_{n}^{\left( a\right) }\left( -z\right) =\left( -\right)
^{n}C_{n}^{\left( a\right) }\left( z\right) $ and $C_{n}^{\left( a\right)
}\left( z\right) $ has $n$ distinct zeros (see, e.g. \cite{abr}), it becomes
clear that $\psi _{+}$ and $\psi _{-}$ have definite and opposite parities,
as expected, and the nodes of $\psi _{+}$ and $\psi _{-}$ just differ by $%
\pm 1$ according to (\ref{10f}). The constants $N_{+}$ and $N_{-}$ are
chosen such that $\int_{-\infty }^{+\infty }dx|\psi _{\pm }|^{2}=|N_{\pm
}|^{2}$ and their absolute values can be determined by substituting (\ref{21}%
) directly into the original first-order coupled equations (\ref{2}) and
demanding a Dirac spinor normalized to unity. By using a couple of
recurrence relations involving the Gegenbauer polynomials (see, e.g. Ref. 
\cite{abr}) one can find that 
\begin{equation}
|N_{\pm }|=\,\sqrt{\frac{E\pm mc^{2}}{2E}}  \label{22}
\end{equation}

Turning now to the isolated solutions, one can observe from (\ref{6}) and (%
\ref{6a}) that a normalizable isolated solution is possible only if the
upper component of the spinor vanishes and $E=-mc^{2}$. The normalized Dirac
spinor can be written as 
\begin{equation}
\psi =\sqrt{\frac{|\gamma |}{\sqrt{\pi }}\,\frac{\Gamma \left( g+1/2\right) 
}{\Gamma \left( g\right) }}\left( 1-z^{2}\right) ^{g/2}\left( 
\begin{array}{l}
0 \\ 
1%
\end{array}
\right)  \label{7a}
\end{equation}

\noindent Note that the lonely hump probability amplitude does exist
independently of the strength of $g$. One can also note that $\mathcal{O}%
\psi =0$ such that there is no state with $E=+mc^{2}$ (for $g>0$).

\section{Conclusions}

We have succeeded in obtaining the complete set of exact bound-state
solutions of fermions in the background of a kink-like potential. Except for
the solution $E=-mc^{2}$, the kink-like potential presents a spectral gap
equal to $2\sqrt{m^{2}c^{4}+\left( \hbar c\gamma \right) ^{2}\left(
2g-1\right) }$. Since $C_{0}^{\left( a\right) }\left( z\right) =1$ (see,
e.g. \cite{abr}) one can see that the position probability amplitude
corresponding to the isolated solution given by (\ref{7a}) can be written in
the very same mathematical structure of the remaining amplitudes. Thus, one
could suspect that the isolated solution is just a particular case and that
its existence is due to the particular method used in this paper. However,
the isolated solution has some distinctive characteristics when compared to
the solutions of the Sturm-Liouville problem which lead us to believe that,
in fact, they belong to a different class of solutions. The isolated
solution breaks the symmetry of the energy levels about $E=0$ exhibited by
the solutions of the Sturm-Liouville problem, and the corresponding
eigenspinor has only one component differing from zero. It is this
asymmetric spectral behaviour that leads to the fractionalization of the
fermion number in quantum field theory \cite{sem}. Furthermore, unlike the
Sturm-Liouville solutions, the isolated solution is there even if the
kink-like potential is not so strong, i.e., there exists an isolated
solution even if $g\leq 1$.

For massless fermions, except for $E=0$, the spectral gap equals to $2\hbar
c|\gamma |\sqrt{2g-1}$ and the Dirac Hamiltonian anticommutes with $\sigma
_{3}$ in such a way that the positive- and negative-eigenenergy solutions
can be mapped by the operation $\psi _{-E}=\sigma _{3}\psi _{E}$. The charge
self-conjugate solution given by (\ref{7a}) appears now in the center of the
spectral gap. As a matter of fact, the kink-like potential used for massless
fermions as a solitonic scalar coupling \cite{jac2} (of course one can not
distinguish a pseudoscalar from a scalar coupling for massless fermions) was
used originally to show the generation of fractional fermion number from the
charge self--conjugate solution.

It is noteworthy that the width of the position probability density for both
class of solutions decreases as $|\gamma |$ or $g$ increases. As such it
promises that the uncertainty in the position can shrink without limit. It
seems that the uncertainty principle fails since such a principle implies
that it is impossible to localize a particle in a region of space less than
half of its Compton wavelength (see, for example, \cite{str}). This apparent
contradiction can be remedied by resorting to the concept of effective
Compton wavelength defined as $\lambda _{eff}=\hbar /(m_{eff}c)$. Hence, the
minimum uncertainty in the position consonant with the uncertainty principle
is given by $\lambda _{eff}/2$ whereas the maximum uncertainty in the
momentum is given by $m_{eff}c$. It means that the localization of a neutral
fermion under the influence of the kink-like potential can shrink to zero
without spoiling the single-particle interpretation of the Dirac equation,
even if the trapped neutral fermion is massless. It is true that as $|\gamma
|$ or $g$ increases the binding potential becomes stronger, though, it
contributes to increase the effective mass of the fermion in such a way that
there is no energy available to produce fermion-antifermion pairs.

As mentioned in the Introduction, the anomalous magnetic-like coupling in
the four-dimensional world turns into a pseudoscalar coupling in the
two-dimensional world. The anomalous magnetic interaction has the form $%
-i\mu \beta \vec{\alpha}\cdot \vec{\triangledown}\phi (r)$, where $\mu $ is
the anomalous magnetic moment in units of the Bohr magneton and $\phi $ is
the electric potential, i.e., the time component of a vector potential \cite%
{tha}. In one-plus-one dimensions the anomalous magnetic interaction turns
into $\sigma _{2}\mu \phi ^{\prime }$, then one might consider the kink
potential as coming from an electric potential proportional to $\ln \left( 
\mathrm{cosh}\,\gamma x\right) ^{g}$. Therefore, the problem addressed in
this paper could be considered as the one of trapping neutral fermions by a
bowl-shaped electric potential.

\bigskip \bigskip \bigskip

\noindent \textbf{Acknowledgments}

The authors wish to thank an anonymous referee for very constructive
remarks. This work was supported in part by means of funds provided by CNPq
and FAPESP.

\end{document}